\documentclass{amsart}

\usepackage{amssymb}

\copyrightinfo{2005}{Evgenij Kritchevski}

\newtheorem{thm}{Theorem}[section]
\newtheorem{prop}[thm]{Proposition}

\numberwithin{equation}{section}

\newcommand{\norm}[1]{\left\Vert#1\right\Vert}
\newcommand{\abs}[1]{\left\vert#1\right\vert}
\newcommand{\set}[1]{\left\{#1\right\}}
\newcommand{\cP}{\mathcal{P}}
\newcommand{\Comp}{\mathbb C}
\newcommand{\Real}{\mathbb R}
\newcommand{\eps}{\varepsilon}
\newcommand{\To}{\rightarrow}
\newcommand{\la}{\lambda}
\newcommand{\of}[1]{\left ( #1 \right ) }
\newcommand{\Hilb}{\mathcal{H}}
\newcommand{\scalp}[2]{\langle#1|#2\rangle}

\begin{document}

\title{Spectral Localization in the Hierarchical Anderson Model}
\author{Evgenij Kritchevski}
\address{Department of Mathematics and Statistics.
McGill University, 805 Sherbrooke Street West Montreal, QC, H3A
2K6 Canada} \curraddr{}

\email{ekritc@math.mcgill.ca}
\thanks{This work was supported in part
by an FQRNT grant.}


\date{Friday, December 16, 2005}

\begin{abstract} We prove that a large class of
hierarchical Anderson models with spectral dimension ${\rm d}\leq
2$ has only pure point spectrum.
\end{abstract}

\maketitle

\section{introduction}

This paper is devoted to study of the spectral properties of the
hierarchical Anderson model and is motivated by the work of
Molchanov \cite{M2}. Before stating our results we recall the
definition of the model and its basic properties. For additional
information about the hierarchical  structures and the
hierarchical Anderson model we refer the reader to \cite{D, BS,
Bo, M1, M2}.

Let $X$ be an infinite countable set.  Throughout the paper
$\delta_x$ will denote the Kronecker delta function at $x\in X$. A
partition $\cP$ of $X$ is a collection of its disjoint subsets
whose union is equal to $X$. Let $\mathbf{n}=(n_r)_{r\geq 0}$ be a
sequence of positive integers and $\mathbf{P}=(\cP_r)_{r\geq 0}$ a
sequence of partitions of $X$. The elements of $\cP_r$ are called
"clusters" of rank $r$.  We say that $(X, \mathbf{P},\mathbf{n})$
is a hierarchical structure if the following hold:
\begin{enumerate}
    \item $n_0=1$ and every $Q\in\cP_0$ has exactly one element.
    \item For $r\geq 1$, every $Q\in \cP_r$ is a  disjoint union of
    $n_r$ clusters in $\cP_{r-1}$.
    \item Given $x,y\in X$, there is a cluster $Q$ of some rank
    containing both $x$ and $y$.

\end{enumerate}
Let us state some immediate consequences of this definition. Every
cluster of rank $r\geq 0$ has size $N_r:=\prod_{s=0}^{r}n_s$.
Given $x\in X$ and $r\geq 0$, there is a unique cluster of rank
$r$ containing $x$. We denote this cluster  by $Q_r(x)$.  The map
 \[ d(x,y):=\min\set{r:y\in
Q_r(x)},\] is a metric on $X$ and  $Q_r(x)=\set{y:d(x,y)\leq r}$.
Note that $Q_r(x)=Q_r(y)$ whenever $d(x,y)\leq r$. Given an integer $n\geq 2$, a hierarchical structure is called \emph{homogeneous of degree} $n$ if $n_r=n$ for all $r\geq 1$.

The free Laplacian on the hierarchical structure $(X,
\mathbf{P},\mathbf{n})$ is defined as follows.
For each $r\geq 0$, let $E_r:l^2(X)\To l^2(X)$ be
the averaging operator
$$(E_r\psi)(x):=\frac{1}{N_r}\sum_{d(x,y)\leq r}\psi(y).$$  Let
 $\mathbf{p}=(p_r)_{r\geq 1}$ be a sequence of positive numbers such that
$\sum_{r=1}^{\infty}p_r=1$.  In the sequel we set $p_0:=0$ and
\[
\lambda_{r}:=\sum_{s=0}^r p_s,\qquad r=0, 1, \cdots, \infty.
\]
The hierarchical Laplacian $\Delta$ on
$l^2(X)$ is defined by
$$\Delta:=\sum_{r=0}^{\infty}p_rE_r.$$
Clearly, $\Delta$ is a bounded self-adjoint
operator and  $0\leq\Delta\leq 1$.

A hierarchical model is a hierarchical structure
$(X,\mathbf{P},\mathbf{n})$ together with the hierarchical
Laplacian $\Delta$. The spectral properties of $\Delta$ only
depend on $\mathbf{n}$ and $\mathbf{p}$ and are summarized in:

\begin{thm}\label{spectral_theorem}
{\rm (1)} The spectrum of $\Delta$  is equal to
    $\{\lambda_r\,:\, r=0, \cdots, \infty\}$.
Each $\lambda_r$, $r<\infty$, is an eigenvalue of $\Delta$ of infinite multiplicity.
The point $\la_\infty=1$ is not
an eigenvalue. \newline
{\rm (2)}  $E_r-E_{r+1}$ is the orthogonal projection onto the
    eigenspace of $\la_r$ and
$$\Delta=\sum_{r=0}^{\infty}\la_r(E_r-E_{r+1}).$$
{\rm (3)} For every $x\in X$, the spectral measure
    for $\delta_x$ and $\Delta$
is given by
$$\mu=\sum_{r=0}^{\infty}\of{\frac{1}{N_r}-\frac{1}{N_{r+1}}}\delta({\la_r}),$$
where $\delta({\la_r})$ stands for the Dirac unit mass at
$\la_r$. Note that $\mu$ does not depend on $x$.
\end{thm}
The spectral measure $\mu$ can be naturally interpreted as the integrated density
of states of the operator  $\Delta$.
Let $x_0\in X$ be given and consider the increasing sequence of
clusters $Q_r(x_0)$, $r\geq 0$. Let  $P_r$ be the orthogonal
projection onto the $N_r$-dimensional subspace
$$l^2(Q_r(x_0)):=\set{\psi\in l^2(X): \psi(x)=0 \textrm{ for }
x\notin Q_r(x_0)}.$$ Let $e_1^{(r)}\leq e_2^{(r)}\leq\dots\leq
e_{N_r}^{(r)}$ be the eigenvalues of the restricted Laplacian $P_r
\Delta P_r$ acting on $l^2(Q_r(x_0))$ and
$$\nu_r:=\frac{1}{N_r}\sum_{s=1}^{r}\delta(e_{s}^{(r)}),$$
the corresponding counting measure.

\begin{prop}\label{densityofstates} The weak-* limit $\lim_{r \rightarrow \infty}\nu_r$ exists and
is equal to $\mu$.
\end{prop}
If
\[\lim_{t\downarrow 0}\frac{\log \mu([1-t,1])} {\log t} ={\rm d}/2,\]
then the number ${\rm d}$ is called the \emph{spectral dimension}
of $\Delta$. This definition is motivated by the analogy with the
edge asymptotics of the density of states of the standard
discrete Laplacian on $\mathbb{Z}^{\rm d}$,  for which the spectral
and spatial dimensions coincide.

The relation $\sum_{y\in
X}\scalp{\delta_x}{\Delta\delta_y}=1$ yields that $\Delta$ generates a
random walk on $X$.
 We recall that the random walk on ${\mathbb Z}^{\rm d}$ generated by the  standard discrete
 Laplacian is recurrent if ${\rm d}=1, 2$ and transient if ${\rm d}> 2$. The corresponding
 result for the hierarchical Laplacian is:
\begin{prop}\label{dimension} Consider a homogeneous
hierarchical structure of degree $n\geq 2$. Suppose that there
exist constants $C_1>0,C_2>0$ and $\rho>1$ such that
\[C_1\rho^{-r}\leq
p_r\leq C_2 \rho^{-r},\]
 for $r$ big enough. Then: \newline {\rm
(1)} The spectral dimension of this model is
   $${\rm d}(n,\rho)=2\frac{\log n}{\log \rho}.$$
    Hence $0<{\rm d}(n,\rho)\leq 2$ iff $ n\leq\rho$. \newline
{\rm (2)} The random walk generated by $\Delta$ is recurrent if
    $0<{\rm d}(n,\rho)\leq 2$ and transient if  ${\rm d}(n,\rho)>2$.
\end{prop}

We now define the hierarchical Anderson model associated to $(X,
{\bf P}, {\bf n})$ and the hierarchical Laplacian $\Delta$.
Consider the probability space $(\Omega, {\mathcal F}, {\mathbb
P})$ where  $\Omega:=\Real^X$,  ${\mathcal F}$ is the  usual Borel
$\sigma$-algebra in $\Omega$, and  ${\mathbb P}$ is a given
probability measure on $(\Omega, {\mathcal F})$. For $\omega \in
\Omega$, we set
$$V_{\omega}:=\sum_{x\in X}\omega(x)\scalp{\delta_x}{\cdot}\delta_x. $$
$V_\omega$ is a self-adjoint (possibly unbounded) multiplication operator on
$l^2(X)$. Let
$$H_{\omega}:=\Delta + V_\omega, \qquad \omega\in\Omega.$$
The family of self-adjoint operators $\{H_\omega\}_{\omega \in
\Omega}$ indexed by the events of the probability space $(\Omega,
{\mathcal F}, {\mathbb P})$ is called the \emph{hierarchical
Anderson model}.

Concerning the probability measure  ${\mathbb P}$, we will need only one technical assumption having to do with the notion of conditional density. Throughout the paper, $m$ will denote the
Lebesgue measure on $\Real$. For any $x\in X$, $\Omega$ can be
decomposed along the $x$'th coordinate as
$\Omega=\Real\times\widetilde{\Omega}$,
$\widetilde{\Omega}=\Real^{X\backslash\set{x}}$. Let
$\widetilde{\mathbb{P}}_x$ be the corresponding  marginal of
$\mathbb{P}$ defined by
$\widetilde{\mathbb{P}}_x(\widetilde{B}):=\mathbb{P}(\Real\times
\widetilde{B})$, where  $\widetilde{B}\subset\widetilde{\Omega}$
is a Borel set. Then for $\widetilde{\mathbb{P}}_x$-a.e.
$\widetilde{\omega}\in\widetilde{\Omega}$, there is a probability
measure $\mathbb{P}^{\widetilde{\omega}}_x$ on $\Real$ s.t. the
conditional Fubini theorem holds: for all $f\in L^1(\Omega,P)$ we
have
$$\int_{\Omega}f(\omega)d\mathbb{P}(\omega)=\int_{\widetilde{\Omega}}\of{\int_{\Real}f(\xi,\widetilde{\omega})d\mathbb{P}^{\widetilde{\omega}}_x(\xi)}d\widetilde{\mathbb{P}}_x(\widetilde{\omega}).$$
If for $\widetilde{\mathbb{P}}_x$-a.e.
$\widetilde{\omega}\in\widetilde{\Omega}$,
$\mathbb{P}^{\widetilde{\omega}}_x$ is absolutely continuous
(a.c.) with respect to $m$, then we say that $\mathbb{P}$ has a
conditional density along the $x$'th coordinate. $\mathbb{P}$ is
called conditionally a.c. if for every $x\in X$, $\mathbb{P}$ has
a conditional density along the $x$'th coordinate. An important
special case of a conditionally a.c. probability measure is the
product measure $\mathbb{P}=\otimes_{x\in X}\mathbb{P}_x$, where
each $\mathbb{P}_x$ is a probability measure on $\Real$ a.c. with
respect to $m$.

We denote by $\sigma_{\rm ac}(H_\omega)$ the absolutely continuous
part of the spectrum of $H_\omega$ and by $\sigma_{\rm
cont}(H_\omega)$ the continuous part. Our main result is:
\begin{thm}\label{maintheorem} Assume that there exists a sequence
$u_r>0$ such that  $\sum_{r=1}^{\infty}u_r^{-1}<\infty$ and
\begin{equation}\label{main_hypothesis}
\sum_{r=1}^{\infty}p_r N_{r-1}u_{r-1}u_r<\infty.
\end{equation}
Then: \newline {\rm (1)} For all $\omega \in \Omega$, $\sigma_{\rm
ac}(H_\omega)=\emptyset$. \newline {\rm (2)} If $\mathbb{P}$ is
conditionally a.c. then $\sigma_{\rm cont}(H_\omega)=\emptyset$
for ${\mathbb P}$-a.e. $\omega$.
\end{thm}
\noindent{\bf Remark 1.}
 Theorem \ref{maintheorem} and Proposition
\ref{dimension} allow to construct hierarchical models with
spectral dimension ${\rm d}\leq 2$ that exhibit Anderson
localization at arbitrary disorder. If $(X,
\mathbf{P},\mathbf{n})$ is a homogeneous hierarchical structure of
degree $n\geq 2$ and $p_r=C \rho^{-r}$ with $\rho>n$, then the
hypothesis \eqref{main_hypothesis} is fulfilled for
$u_r=r^{1+\eps}$. Given $0<{\rm d}<2$, one can adjust $\rho>n$ to
make ${\rm d}(n,\rho)={\rm d}$. If $p_r=Cr^{-3-\eps}n^{-r}$, then
the model has spectral dimension ${\rm d}=2$ and
\eqref{main_hypothesis} is verified for $u_r=r^{1+\eps/3}$. One
can also construct trivial models with ${\rm d}=0$ by taking $p_r$
to decrease faster than $\rho^{-r}$ for any $\rho$.  We emphasize
that homogeneity of the hierarchical structure is not required for
Theorem \ref{maintheorem}. \newline \noindent{\bf Remark 2.} In
\cite{M2}, Molchanov  has  proven  that if the random variables
$\omega(x)$ are i.i.d. with a Cauchy distribution, then  Theorem
\ref{maintheorem} holds under the condition
$$\sum_{r=1}^{\infty}p_ru_r<\infty.$$
In particular, in this case the theorem holds for $\Delta$  of any
spectral dimension. Molchanov's argument is based on subtle
properties of Cauchy random variables and cannot be directly
extended to any other probability measure. In contrast, our proof
of localization in spectral dimension $\mathrm{d}\leq 2$ is based
on general arguments and is the first step in extending
Molchanov's result to  a more general class of probability
measures. \newline \noindent {\bf Remark 3.} The fractional
moments method of Aizenman and Molchanov \cite{AM} allows to prove
localization for $\Delta + \sigma V_\omega$ for large disorder
$\sigma$ or for large energies. One needs an extra decoupling
hyphothesis on the random variables $\omega(x)$ and the condition
on $\Delta$ that
\begin{equation}\label{AizMolcondition}
B:=\sup_{x}\sum_{y\in
X}\abs{\scalp{\delta_x}{\Delta\delta_y}}^s<\infty
\end{equation}
for some $0<s<1$. Simple estimates show that
$$\sum_{r=1}^{\infty}p_rN_r^{1-s}\leq B \leq \sum_{r=1}^{\infty} p_r^s N_r^{1-s}.$$
The requirement \eqref{AizMolcondition} on the decay of $p_r$ is
comparable to the hypothesis \eqref{main_hypothesis}, while
Theorem \ref{maintheorem} is valid at arbitrary disorder or
energy. \newline
\noindent {\bf Remark 4.}  Part (2) of Theorem \ref{maintheorem} does not hold
for all $\omega$.  Our method of proof combined with  the general results of \cite{DMS},
\cite{G}
 yields that $H_\omega$ will have singular continuous spectrum for some $\omega$'s.

\section{The free Laplacian}
In this section, we prove Theorem \ref{spectral_theorem},
Proposition \ref{densityofstates} and Proposition \ref{dimension}.

\emph{Proof of Theorem \ref{spectral_theorem}.} For $r\geq 0$, let
$\Hilb_r=Ran(E_r)$. $\Hilb_r$ is the closed subspace of $l^2(X)$
consisting of functions that are constant on each cluster of rank
$r$. Note that
$$l^2(X)=\Hilb_0\supset \Hilb_1\supset\Hilb_{2}\supset\Hilb_{3}\supset\dots $$
and that $\bigcap{\Hilb_r}=\set{0}$ since a nonzero function
constant on every cluster would have infinite $l^2$ norm. These
observations yield that
\begin{equation}\label{eigenspaces}
    l^2(X)=\bigoplus_{r=0}^{\infty} L_r,
\end{equation}
where $L_r$ is the orthogonal complement of $\Hilb_{r+1}$ in
$\Hilb_r$. Note that $L_r$ is the infinite dimensional subspace of
functions $\psi$ s.t. $E_s\psi=\psi$ for $0\leq s \leq r$ and
$E_s\psi=0$ for $s>r$. Hence for every $\psi\in L_r$,
$\Delta\psi=\la_r\psi$, and this proves parts (1) and (2).

The spectral measure $\mu_{x,\Delta}$ for $\delta_x$ and $\Delta$
is the unique Borel probability measure on $\Real$ s.t.
$$\scalp{\delta_x}{f(\Delta)\delta_x}=\int_{\Real}f(\xi)d\mu_{x,\Delta}(\xi),$$
for every bounded Borel function $f:\Real\To\Comp$. To compute
$\mu_{x,\Delta}$, we decompose $\delta_x$ according
to \eqref{eigenspaces}:
$$\delta_x=\sum_{r=0}^{\infty}(E_r-E_{r+1})\delta_x=\sum_{r=0}^{\infty}\of{\frac{1}{N_r}\mathbf{1}_{Q_r(x)}- \frac{1}{N_{r+1}}\mathbf{1}_{Q_{r+1}(x)} },$$
where $\mathbf{1}_{Q_r(x)}:=\sum_{y\in Q_r(x)}\delta_y$.
Hence
$$f(\Delta)\delta_x= \sum_{r=0}^{\infty}f(\la_r)\of{\frac{1}{N_r}\mathbf{1}_{Q_r(x)}- \frac{1}{N_{r+1}}\mathbf{1}_{Q_{r+1}(x)} },$$
and
$$\scalp{\delta_x}{f(\Delta)\delta_x}=\sum_{r=0}^{\infty}f(\la_r)\norm{\frac{1}{N_r}\mathbf{1}_{Q_r(x)}- \frac{1}{N_{r+1}}\mathbf{1}_{Q_{r+1}(x)} }^2.$$
Since $\norm{\frac{1}{N_r}\mathbf{1}_{Q_r(x)}-
\frac{1}{N_{r+1}}\mathbf{1}_{Q_{r+1}(x)} }^2=1/N_r-1/N_{r+1}$, (3)
follows. $\Box$

The analysis of the density of states of $\Delta$ is facilitated if one introduces
the cut-off Laplacians
$$\Delta_r:=\sum_{s=0}^{r}p_s E_s, \qquad r\geq 0.$$
It is technically easier to work with $\Delta_r$ than with
$P_r\Delta P_r$. Note that $l^2(Q_r(x_0))$ is an invariant
subspace for $\Delta_r$. One can exactly compute the eigenvalues
and eigenvectors of the restricted operator $P_r\Delta_r$ acting
on $l^2(Q_r(x_0))$. If $0\leq s\leq r$, then every $\psi\in
L_s\cap l^2(Q_r(x_0))$ is an eigenvector of $P_r\Delta_r$ with
eigenvalue $\la_r$. The subspace $L_s\cap l^2(Q_r(x_0))$ has
dimension $D_s^{(r)}:=N_r(1/N_s-1/N_{s+1})$ for $0\leq s\leq r-1$,
and the subspace $L_r\cap l^2(Q_r(x_0))$ has dimension
$D_r^{(r)}:=1$. Since $\sum_{s=0}^rD_s^{(r)}=N_r$, the spectrum of
$P_r\Delta_r$ is equal to $\{\lambda_s\,:\, s=0, \cdots, r\}$ and
each eigenvalue $\la_s$ has multiplicity $D_s^{(r)}$.

\emph{Proof of Proposition \ref{dimension}} . Let $\nu^*$ be a
weak-* limit point of the sequence $\nu_r$. Let $\nu_{r_k}$ be a
subsequence converging to $\nu^{*}$. We claim that
\begin{equation}\label{nustar}
\nu^*({\set{\la_s}})=\mu(\set{\la_s}),
\end{equation}
for all $s\geq 0$. Indeed, let $\delta:=\min_{j\neq
s}\abs{\la_s-\la_j}/2$ and $0<\eps<\delta/3$. Since $\norm{P_r\Delta P_r -
P_r\Delta_r}\leq \sum_{j=r+1}^{\infty}p_j$, we have that
$\norm{P_r\Delta P_r - P_r\Delta_r}\leq\eps$ for all $r$ big
enough. For such $r$, the spectrum of $P_r\Delta P_r$ is contained in $\bigcup_{j=0}^{r}[\la_j-\eps,\la_j+\eps]$.
Let $R$ be the spectral projection of $P_r\Delta P_r$ on $[\la_s-\eps,\la_s+\eps]$ and $T$ the spectral projection of
$P_r\Delta_r$ on the same interval. Let $\gamma$ be the circle $\set{z\in\Comp:\abs{z-\la_s}=\delta}$, oriented counterclockwise. Then
\begin{equation*}
\begin{split}
R-T&=\frac{1}{2\pi i}\oint_\gamma (z-P_r\Delta P_r)^{-1}dz-\frac{1}{2\pi i}\oint_\gamma (z-P_r\Delta_r)^{-1}dz\\
&=\frac{1}{2\pi i}\oint_\gamma (z-P_r\Delta P_r)^{-1}(P_r\Delta P_r-P_r\Delta_r)(z-P_r\Delta_r)^{-1}dz,
\end{split}
\end{equation*}
and thus
$$\norm{R-T}\leq \delta (2\delta/3)^{-1}\eps (2\delta/3)^{-1}\leq 3/4<1.$$
It follows that $Ran(R)$ and $Ran(T)$ have the same dimension and that
$$\#\set{s:e_s^{(r)}\in [\la_s-\eps,\la_s+\eps]}=D_s^{(r)}Ž
.$$ Then for all $k$ big enough
$$\nu_{r_k}([\la_s-\eps,\la_s+\eps])=D_s^{(r)}/N_r=1/N_s -1/N_{s+1}.$$
Letting $k\to\infty$, we get
$\nu^*([\la_s-\eps,\la_s+\eps])=1/N_s -1/N_{s+1}$, and \eqref{nustar}
follows by taking $\eps \downarrow 0$. Since
$\sum_{s=0}^\infty(1/N_s-1/N_{s+1})=1$ and $\nu^*$ is a
probability measure, we must have that $\nu^*=\mu$. Therefore
$\mu$ is the unique weak-* limit point of the sequence $\nu_r$ and
$\lim_{r\To\infty}\nu_r=\mu$. $\Box$

\emph{Proof of Proposition \ref{dimension}}. Note that
$\mu([1-t,1])$ is a piecewise constant function of $t$ with jump discontinuities at
the points $1-\la_r$. Since
$$C_1(\rho-1)^{-1}\rho^{-r}\leq  1-\la_r=\sum_{s=r+1}^{\infty}p_s\leq C_2(\rho-1)^{-1}\rho^{-r},$$
and $\mu([1-\la_r,1])=1/N_r=n^{-r}$, we have that
$$\lim_{t\downarrow 0}\frac{\log \mu([1-t,1])}{\log t}=\frac{\log n}{\log \rho},$$
which proves (1).

The random walk on $X$ starting at $x$ is transient if
$R:=\sum_{k=0}^{\infty}\scalp{\delta_x}{\Delta^k \delta_x}<\infty$
and recurrent if $R=\infty$. Part (3) of Theorem
\ref{spectral_theorem} allows to compute $R$ explicitly:
$$R= \scalp{\delta_x}{(1-\Delta)^{-1} \delta_x}= \int \frac{d\mu(\xi)}{1-\xi}
=\sum_{r=0}^{\infty}\frac{N_r^{-1}-N_{r+1}^{-1}}{1-\la_r}.$$
The bounds
$$C_2^{-1}(\rho-1)(1-1/n)\sum_{r=0}^{\infty}(\rho/n)^r\leq R\leq C_1^{-1}(\rho-1)(1-1/n)\sum_{r=0}^{\infty}(\rho/n)^r$$
show that $R<\infty$ for $\rho<n$ and $R=\infty$ for $\rho\geq n$,
and part (2) follows. $\Box$

\section{Proof of the localization theorem}
This section is devoted to the proof of Theorem \ref{maintheorem}
and is organized as follows. We first derive a hierarchical
approximation formula for the resolvent $(H_\omega-z)^{-1}$. Then
we use the formula to obtain a bound on the resolvent matrix
elements. This bound combined with the Simon-Wolff localization
criterion yields the statement.

Set
$$H_{\omega,r}:=V_{\omega} +\sum_{s=0}^{r}p_sE_s, \qquad r \geq 0.$$
Fix $\omega\in\Omega$. For any $Q_r\in\cP_r$, the subspace
$l^2(Q_r)$ is invariant for $H_{\omega,r}$. Let
$\sigma(\omega,Q_r)$ be the set of the eigenvalues of the
restricted operator $H_{\omega,r}\upharpoonright l^2(Q_r)$ and
$\sigma_{\omega}:=\bigcup \sigma(\omega,Q_r)$ where the union is
over all clusters of all ranks. Clearly,  $\sigma_{\omega}$ is a
countable subset of $\Real$. For
$z\in\Comp\backslash\sigma_{\omega}$, $r\geq 0$, and $x,y\in X$,
we set
$$G_{\omega,r}(x,y;z):=\scalp{\delta_x}{(H_{\omega,r}-z)^{-1}\delta_y}.$$
For $z\in\Comp\backslash\sigma_{\omega}$, $r\geq 0$ and $t\in X$,
let $g_{\omega,r}(t;z)$ be the average of
$G_{\omega,r}(\cdot,t;z)$ over the cluster $Q_r(t)$, i.e.
$$g_{\omega,r}(t;z):=\frac{1}{N_r}\sum_{d(t',t)\leq r}G_{\omega,r}(t',t;z).$$
Since the joint spectral measure for $\delta_t$, $\delta_{t'}$ and
$H_{\omega,r}$ is real,
$G_{\omega,r}(t',t;z)=G_{\omega,r}(t,t';z)$ and
\begin{equation}\label{symmetry_of_G}
g_{\omega,r}(t;z)=\frac{1}{N_r}\sum_{d(t',t)\leq
r}G_{\omega,r}(t,t';z)=\frac{1}{N_r}\scalp{\delta_t}{(H_{\omega,r}-z)^{-1}\mathbf{1}_{Q_r(t)}}.
\end{equation}
\begin{prop}\label{expansion_of_resolvent}
Let $\omega\in\Omega$, $x,y\in X$,
$z\in\Comp\backslash\sigma_{\omega}$ and $r\geq 0$ be given. Then
\begin{equation}\label{expansion_formula}
G_{\omega,r}(x,y;z)=G_{\omega,0}(x,y;z)-\sum_{s=d(x,y)}^{r}p_sN_{s-1}g_{\omega,s-1}(x;z)g_{\omega,s}(y;z).
\end{equation}
\end{prop}
\begin{proof} The formula holds for $r=0$ since $p_0=0$.  For $s\geq 1$, the resolvent identity yields
$$(H_{\omega,s}-z)^{-1}\delta_y-(H_{\omega,s-1}-z)^{-1}\delta_y=-(H_{\omega,
s-1}-z)^{-1}p_s E_s(H_{\omega,s}-z)^{-1}\delta_y.$$ Observe that
$E_s(H_{\omega,s}-z)^{-1}\delta_y
=g_{\omega,s}(y;z)\mathbf{1}_{Q_s(y)}$. Taking
$\scalp{\delta_x}{\cdot}$ in the above equation yields
\begin{equation}\label{interation}
G_{\omega,s}(x,y;z)-G_{\omega,s-1}(x,y;z)=-p_s
g_{\omega,s}(y;z)\scalp{\delta_x}{(H_{\omega,s-1}-z)^{-1}\mathbf{1}_{Q_s(y)}}.
\end{equation}
Note that by \eqref{symmetry_of_G}, $$\scalp{\delta_x}{(H_{\omega,s-1}-z)^{-1}\mathbf{1}_{Q_s(y)}}=\left\{%
\begin{array}{ll}
    N_{s-1} g_{\omega,s-1}(x;z), & \hbox{if } d(x,y)\leq s, \\
    0, & \hbox{if } d(x,y)>s. \\
\end{array}%
\right.$$ The formula \eqref{expansion_formula} follows after
adding \eqref{interation} for $s=1,2,\cdots,r$.
\end{proof}
The key step in our proof is:
\begin{thm}\label{bound_for_G} Suppose that $p_r$ and $N_r$
satisfy \eqref{main_hypothesis}. Let $\omega\in\Omega$ and $x\in
X$ be fixed. Then for $m$-a.e. $e\in \Real\backslash
\sigma_\omega$,
\begin{equation}\label{eq_bound_for_G}
\sup_{r\geq 0}\sum_{y\in X}\abs{G_{\omega,r}(x,y;e)}^2<\infty.
\end{equation}
\end{thm}
\begin{proof} We shall use the
following general result, proven in \cite{M2}:\newline Let $A$ be a
hermitian $N\times N$ matrix and $v\in\Comp^N$. Then for all
$M>0$,
\begin{equation}\label{general_lemma}
m\of{\set{e: \norm{(A-e)^{-1}v}_2^2\geq M}}\leq
4\sqrt{\frac{N}{M}}\norm{v}_2,
\end{equation} where $\norm{\cdot}_2$  stands for
the $l^2$ norm on $\Comp^N$.

Since $l^2(Q_r(x))$ is an $N_r$-dimensional invariant subspace for
$H_{\omega,r}$ and since
$\norm{\mathbf{1}_{Q_r(x)}}_2=\sqrt{N_r}$, we have from
\eqref{general_lemma} that for $M_r>0$,

$$m\of{\set{e\in\Real\backslash\sigma_{\omega}:\norm{(H_{\omega,r}-e)^{-1}\mathbf{1}_{Q_r(x)}}_2^2\geq M_r}}\leq \frac{4N_r}{\sqrt{M_r}}.$$
Let  $M_r>0$ be  a sequence satisfying
$\sum_{r=1}^{\infty}N_rM_r^{-1/2}<\infty$. By the Borel-Cantelli lemma,
for $m$-a.e.
$e\in\Real\backslash\sigma_{\omega}$, there exists a finite
constant $C_e$ such that
\begin{equation}\label{bound_from_BC}
\norm{(H_{\omega,r}-e)^{-1}\mathbf{1}_{Q_r(x)}}_2^2<C_e M_r,
\end{equation}
for all $r\geq 0$. From now on, such an
$e\in\Real\backslash\sigma_{\omega}$ is fixed. Using the
representation formula \eqref{expansion_formula}, we get the
estimate
\begin{equation}\label{cauchy_schwarz}
\begin{split}
\of{\sum_{y\in X}\abs{G_{\omega,r}(x,y;e)}^2}^{1/2}& \leq \abs{G_{\omega,0}(x,x;e)}\\
&+\sum_{s=1}^{r}p_sN_{s-1}\abs{g_{\omega,s-1}(x;e)}\of{\sum_{d(x,y)\leq
s}\abs{g_{\omega,s}(y;e)}^2}^{1/2}.\\
\end{split}
\end{equation}
Observe that
\begin{equation*}
\begin{split}
\of{\sum_{d(x,y)\leq s}\abs{g_{\omega,s}(y;e)}^2}^{1/2} &=
\of{\sum_{d(x,y)\leq
s}\abs{\frac{1}{N_s}\scalp{\delta_y}{(H_{\omega,s}-e)^{-1}\mathbf{1}_{Q_s(y)}}}^2}^{1/2}
\\
&=\frac{1}{N_s}\of{\sum_{d(x,y)\leq
s}\abs{\scalp{\delta_y}{(H_{\omega,s}-e)^{-1}\mathbf{1}_{Q_s(x)}}}^2}^{1/2}
\\
&= \frac{1}{N_s}\norm{(H_{\omega,s}-e)^{-1}\mathbf{1}_{Q_s(x)}}_2.
\end{split}
\end{equation*}
Inequality \eqref{bound_from_BC} gives the bound
\begin{equation}\label{bound_for_gry}
\of{\sum_{d(x,y)\leq s}\abs{g_{\omega,s}(y;e)}^2}^{1/2}\leq
C_e^{1/2}\frac{\sqrt{M_s}}{N_s}.
\end{equation}
Moreover
\begin{equation}\label{bound_for_grx}
N_{s-1}\abs{g_{\omega,s-1}(x;e)}=\abs{\scalp{\delta_x}{(H_{\omega,s-1}-e)^{-1}\mathbf{1}_{Q_{s-1}(x)}}}\leq
C_e^{1/2}\sqrt{M_{s-1}}.
\end{equation}
Combination of \eqref{cauchy_schwarz} with \eqref{bound_for_grx}
and \eqref{bound_for_gry}  yields the estimate
$$\of{\sum_{y\in X}\abs{G_{\omega,r}(x,y;e)}^2}^{1/2}\leq \abs{G_{\omega,0}(x,x;e)} + C_e\sum_{s=1}^{r}p_s\frac{\sqrt{M_s}{\sqrt{M_{s-1}}}}{N_s}.$$
By hypothesis \eqref{main_hypothesis}, the sequence $M_r=(u_r
N_r)^2$ satisfies
$$\sum_{r=1}^{\infty}N_rM_r^{-1/2}=\sum_{r=1}^{\infty}u_r^{-1}<\infty.$$
Since
$$\sum_{r=1}^{\infty}p_r\frac{\sqrt{M_r}{\sqrt{M_{r-1}}}}{N_r}=\sum_{r=1}^{\infty}p_r N_{r-1}u_{r-1}u_r <\infty,$$
the result follows.
\end{proof}

Let us recall the Simon-Wolff localization criterion. For $x\in X$
and $\omega\in\Omega$, denote by $\mu_x^{\omega}$ the spectral
measure for $\Delta+V_\omega$ and $\delta_x$, by
$\mu_{x,\mathrm{cont}}^{\omega}$ the continuous part of
$\mu_x^{\omega}$ and by $\mu_{x,\mathrm{ac}}^{\omega}$ the a.c.
part. Define the function $G_{\omega,x}:\Real\To [0,+\infty]$ by
$$G_{\omega,x}(e):=\int_{\Real}\frac{d \mu_x^{\omega}(\lambda)}{(e-\la)^2}
=\lim_{\epsilon\downarrow
0}\norm{(\Delta+V_\omega-e-i\epsilon)^{-1}\delta_x}^2.$$ By the
Theorem of de la Vall\'{e} Poussin,
$$d\mu_{x,\mathrm{ac}}^{\omega}(e)=\pi^{-1}\of{\lim_{\epsilon\downarrow
0}\eps\norm{(\Delta+V_\omega-e-i\epsilon)^{-1}\delta_x}^2}de.$$
Hence, if for a fixed $\omega\in\Omega$ we have that
$G_{\omega,x}(e)<\infty$ for $m$-a.e. $e\in\Real$, then
$\mu_{x,\mathrm{ac}}^{\omega}=0$.

The Simon-Wolff localization criterion is summarized in:
\begin{thm}\label{SWthm}  Assume
that $\mathbb{P}$ has a conditional density along the $x$'th
coordinate. Let $B\subset\Real$ be a Borel set such that
$G_{\omega,x}(e)<\infty$ for $\mathbb{P}\otimes m$-a.e.
$(\omega,e)\in\Omega\times B$. Then
$\mu_{x,\mathrm{cont}}^{\omega}(B)=0$ for $\mathbb{P}$-a.e.
$\omega\in\Omega$.
\end{thm}
Theorem \ref{SWthm} is a well known consequence of the rank-1
Simon-Wolff theorem \cite{SW} and the conditional Fubini theorem.

\emph{Proof of Theorem \ref{maintheorem}.} Fix $\omega\in\Omega$
and fix $e\in\Real\backslash\sigma_{\omega}$ for which the bound
\eqref{eq_bound_for_G} holds. By monotone convergence
$$ \int_{\Real}\frac{d \mu_x^{\omega}(\lambda)}{(e-\la)^2}=
\lim_{\eps\downarrow 0} \int_{\Real}\frac{d
\mu_x^{\omega}(\lambda)}{(e-\la)^2+\eps^2}=
\sup_{\eps>0}\int_{\Real}\frac{d
\mu_x^{\omega}(\lambda)}{(e-\la)^2+\eps^2}.$$ Since for any
$z\in\Comp\backslash\Real$,
$$\lim_{r\To\infty}\norm{(H_{\omega,r}-z)^{-1}-
(H_{\omega}-z)^{-1}}=0,$$ we have that the weak-* limit
$\lim_{r\To\infty}\mu_{x,r}^{\omega}$ equals $\mu_x^{\omega}$, where
$\mu_{x,r}^{\omega}$ is the spectral measure for $H_{\omega,r}$
and $\delta_x$. Therefore
\begin{equation*}
\begin{split}\int_{\Real}\frac{d \mu_x^{\omega}(\lambda)}{(e-\la)^2}=
\sup_{\eps>0}\lim_{r\To\infty}\int_{\Real}\frac{d
\mu_x^{\omega,r}(\lambda)}{(e-\la)^2+\eps^2} &\leq \sup_{\eps>0,r\geq1}\int_{\Real}\frac{d
\mu_x^{\omega,r}(\lambda)}{(e-\la)^2+\eps^2}\\
&=\sup_{r\geq1}\int_{\Real}\frac{d
\mu_x^{\omega,r}(\lambda)}{(e-\la)^2}\\
&=\sup_{r\geq 1}\norm{(H_{\omega,r}-e)^{-1}\delta_x}^2\\
&=\sup_{r\geq 1}\sum_{y\in X}\abs{G_{\omega,r}(x,y)}^2<\infty.
\end{split}
\end{equation*}
In the final equality we used the fact that $\set{\delta_y:y\in
X}$ is an orthonormal basis for $l^2(X)$. Since
$m\of{\sigma_\omega}=0$ and since the bound \eqref{eq_bound_for_G}
holds for $m$-a.e. $e\in\Real\backslash\sigma_\omega$, we have
that for every fixed $\omega\in\Omega$, $G_{\omega,x}(e)<\infty$ for $m$-a.e. $e\in\Real$. This proves part (1). Part (2) follows from the fact that $G_{\omega,x}(e)<\infty$ for $\mathbb{P}\otimes m$-a.e.
$(\omega,e)\in\Omega\times \Real$ and the Simon-Wolff criterion.  $\Box$

\textbf{Acknowledgements.} We are grateful to Professor Vojkan
Jaksic, who suggested this research project, and from whom the
author learned about random Schr\"{o}dinger operators. We would
like to thank Professor Stanislav Molchanov for very helpful
discussions and encouragements during the author's visit at the
University of Charlotte, NC. We also benefited from discussions
with the following people: Kingwood Chen, Serguei Denissov, Marco
Merkli, Juan-Manuel Perez-Abarca and Nicola Squartini. A very
special thank goes to Kingwood Chen for hospitality during the
author's visit at UNCC.

\end{document}